\documentstyle[12pt]{article}
\newcommand\bq{\openup-2\jot\begin{quotation}}
\newcommand\eq{\end{quotation}\openup2\jot}

\begin{document}
\title{Quantum Philosophy:\\ The Flight from Reason in Science}

\author{Sheldon Goldstein\\
Department of Mathematics, Rutgers University\\ New Brunswick, NJ
08903}
\date{January 10, 1996}
\maketitle
\openup2\jot
I want to discuss a rather delicate matter concerning a notoriously
difficult subject, the foundations of quantum mechanics, a subject that
has inspired a great many peculiar proclamations. Some examples:

\bq
\noindent \dots\ the idea of an objective real world whose smallest parts exist
objectively in the same sense as stones or trees exist, independently of
whether or not we observe them \dots\  is impossible \dots\  \cite{Heisenberg1}
\eq
and
\bq We can no longer speak of the behavior of the particle independently of
the process of observation. As a final consequence, the natural laws
formulated mathematically in quantum theory no longer deal with the
elementary particles themselves but with our knowledge of them. Nor is it
any longer possible to ask whether or not these particles exist in space
and time objectively \dots\   	 	

\dots\ Science no longer confronts nature as an objective observer, but
sees itself as an actor in this interplay between man and nature. The
scientific method of analysing, explaining, and classifying has become
conscious of its limitations \dots\ method and object can no longer be
separated. \cite{Heisenberg2}
\eq 
and 
\bq A complete elucidation of one and the same object may require diverse
points of view which defy a unique description. Indeed, strictly speaking, the
conscious analysis of any concept stands in a relation of exclusion to its
immediate application. \cite{Bohr} 
\eq 
This last quotation is an expression of what has traditionally been called
complement\-arity---but what might nowadays be called multiphysicalism.

For my purposes here, what is most relevant about these sentiments is that
they were expressed, not by lay popularizers of modern science, nor by its
postmodern critics, but by Werner Heisenberg and Niels Bohr, the two
physicists most responsible, with the possible exception of Erwin
Schr\"odinger, for the creation of quantum theory. It does not require
great imagination to suggest that there is little in these sentiments with
which a postmodernist would be inclined to disagree and much that he or she
would be happy to regard as compelling support for the postmodern
enterprise (see, for example, \cite{pl,aro,fk}).

The ``quantum philosophy'' expressed by such statements is part of the
Copenhagen interpretation of quantum theory, which, in addition to the
vagueness and subjectivity suggested by the preceding quotes, also
incorporated as a central ingredient the notion that in the microscopic
quantum domain the laws of nature involve irreducible randomness.  The
Copenhagen interpretation was widely, I would say at one time almost
universally, accepted within the physics community, though there were some
notable exceptions, such as Einstein and Schr\"odinger.  Here is
Schr\"odinger in 1926 \cite[page 228]{Schr1}:

\bq Bohr's \dots\   approach to atomic problems \dots\  is really
remarkable. He is completely convinced that any understanding in the usual
sense of the word is impossible. Therefore the conversation is almost
immediately driven into philosophical questions, and soon you no longer
know whether you really take the position he is attacking, or whether you
really must attack the position he is defending.
\eq
and Schr\"odinger in 1959 \cite[page 472]{Schr2}:
\bq
With very few exceptions (such as Einstein and Laue) all the rest of the
theoretical physicists were unadulterated asses and I was the only sane
person left. \dots\ The one great dilemma that ails us \dots\ day and night
is the wave-particle dilemma. In the last decade I have written quite a lot
about it and have almost tired of doing so: just in my case the effect is
null \dots\ because most of my friendly (truly friendly) nearer colleagues
(\dots\ theoretical physicists) \dots\ have formed the opinion that I
am---naturally enough---in love with `my' great success in life (viz., wave
mechanics) reaped at the time I still had all my wits at my command and
therefore, so they say, I insist upon the view that `all is waves'. Old-age
dotage closes my eyes towards the marvelous discovery of `complementarity'.
So unable is the good average theoretical physicist to believe that any
sound person could refuse to accept the Kopenhagen oracle\dots
\eq

Einstein in 1949 \cite{Einstein} offered a somewhat more constructive response:
\bq I am, in fact, rather firmly convinced that the essentially statistical
character of contemporary quantum theory is solely to be ascribed to the
fact that this (theory) operates with an incomplete description of physical
systems \dots\ 

[In] a complete physical description, the statistical quantum theory would
\dots\  take an approximately analogous position to the statistical mechanics
within the framework of classical mechanics \dots\ 
\eq
Part of what Einstein is saying here is that (much of) the apparent
peculiarity of quantum theory, and in particular its randomness, arises
from mistaking an incomplete description for a complete one.

In view of the radical character of quantum philosophy, the arguments
offered in support of it have been surprisingly weak. More remarkable still
is the fact that it is not at all unusual, when it comes to quantum
philosophy, to find the very best physicists and mathematicians making
sharp emphatic claims, almost of a mathematical character, that are
trivially false and profoundly ignorant. For example, John von Neumann, one
of the greatest mathematicians of this century, claimed to have
mathematically proven that Einstein's dream, of a deterministic completion
or reinterpretation of quantum theory, was mathematically impossible. He
concluded that \cite{vN}  

\bq It is therefore not, as is often assumed, a question of a
re-interpretation of quantum mechanics---the present system of quantum
mechanics would have to be objectively false, in order that another
description of the elementary processes than the statistical one be
possible.
\eq

This claim of von Neumann was, of course, just about universally accepted.
For example, Max Born, who formulated the statistical interpretation of the
wave function, assures us that \cite{born}

\bq No concealed parameters can be introduced with the help of which the
indeterministic description could be transformed into a deterministic one.
Hence if a future theory should be deterministic, it cannot be a
modification of the present one but must be essentially different. (Born 1949)
\eq

However, in 1952 David Bohm, through a refinement of de Broglie's pilot
wave model of 1927, found just such a reformulation of quantum
theory\cite{Bohm52}.  Bohm's theory, Bohmian mechanics, was precise,
objective, and deterministic---not at all congenial to quantum philosophy
and a counterexample to the claims of von Neumann. Nonetheless, we still
find, more than a quarter of a century after the discovery of Bohmian
mechanics, statements such as these:

\bq The proof he [von Neumann] published \dots, though it was made much
more convincing later on by Kochen and Specker, still uses assumptions
which, in my opinion, can quite reasonably be questioned. \dots\ In my
opinion, the most convincing argument against the theory of hidden
variables was presented by J. S. Bell (1964). (Eugene Wigner 1976
\cite{Wigner1})
\eq 
and
\bq This [hidden variables] is an interesting idea and even though few of
us were ready to accept it, it must be admitted that the truly telling
argument against it was produced as late as 1965, by J. S.  Bell. \dots\  This
appears to give a convincing argument against the hidden variables theory.
(Wigner 1983 \cite{Wigner2})
\eq

Now there are many more statements of a similar character that I could have
cited; I chose these partly because Wigner was not only one of the leading
physicists of his generation, but, unlike most of his contemporaries, he
was also profoundly concerned with the conceptual foundations of quantum
mechanics and wrote on the subject with great clarity and insight.

There was, however, one physicist who wrote on this subject with even greater
clarity and insight than Wigner himself, namely the very J. S. Bell whom
Wigner praises for demonstrating the impossibility of a deterministic
completion of quantum theory such as Bohmian mechanics. So let's see how
Bell himself reacted to Bohm's discovery: 

\bq But in 1952 I saw the impossible done.  It was in papers by David
Bohm. Bohm showed explicitly how parameters could indeed be introduced,
into nonrelativistic wave mechanics, with the help of which the
indeterministic description could be transformed into a deterministic one.
More importantly, in my opinion, the subjectivity of the orthodox version,
the necessary reference to the `observer,' could be eliminated. \cite[page
160]{Bell}
\eq
and Bell again

\bq Bohm's 1952 papers on quantum mechanics were for me a
revelation. The elimination of indeterminism was very striking. But more
important, it seemed to me, was the elimination of any need for a vague
division of the world into ``system'' on the one hand, and ``apparatus'' or
``observer'' on the other. I have always felt since that people who have
not grasped the ideas of those papers \dots\ and unfortunately they remain the
majority \dots\ are handicapped in any discussion of the meaning of quantum
mechanics. \cite[page 173]{Bell}
\eq

Wigner to the contrary notwithstanding, Bell did not establish the
impossibility of a deterministic reformulation of quantum theory, nor did
he ever claim to have done so. On the contrary, over the course of the past
several decades, until his untimely death several years ago, Bell was the
prime proponent, for a good part of this period almost the sole proponent,
of the very theory, Bohmian mechanics, that he is supposed to have
demolished. What Bell did demonstrate is the remarkable conclusion that
nature, if governed by the predictions of quantum theory, must be nonlocal,
exhibiting surprising connections between distant events. And unlike the
claims of quantum philosophy, this nonlocality {\it is\/} well founded,
and, with the experiments of Aspect \cite{Aspect}, rather firmly
established. Nonetheless, {\it it\/} is far from universally accepted by
the physics community. Here is Bell, expressing his frustration at the
obtuseness of his critics, and insisting that his argument for nonlocality
involves no unwarranted assumptions:
 
\bq\noindent Despite my insistence that the determinism was inferred rather
than assumed, you might still suspect somehow that it is a preoccupation
with determinism that creates the problem. Note well then that the
following argument makes no mention whatever of determinism. ...  Finally
you might suspect that the very notion of particle, and particle orbit
\dots\ has somehow led us astray. \dots\ So the following argument will not
mention particles \dots\  nor any other picture of what goes on at the
microscopic level. Nor will it involve any use of the words `quantum
mechanical system', which can have an unfortunate effect on the discussion.
The difficulty is not created by any such picture or any such terminology.
It is created by the predictions about the correlations in the visible
outputs of certain conceivable experimental set-ups. \cite[page 150]{Bell}
\eq

So what is the relevance of what I've described to the theme of this
conference? Well, there's some bad news and some good news. The bad news,
nothing you didn't already know anyway, is that objectivity is difficult to
maintain and that physicists, even in their capacity as scientists, are
only human. Nothing new.  I must say, however, that the complaceny of the
physics establishment with regard to the foundations of quantum mechanics
has been, it seems to me, somewhat astonishing, though I must admit to
lacking sufficient historical perspective to have genuine confidence that
what has occurred is at all out of the ordinary. But let me once again
quote Bell:

\bq  But why then had Born not told me of this `pilot wave'?  If only
to point out what was wrong with it? Why did von Neumann not consider it?
\dots\  Why is the pilot wave picture ignored in text books?  Should it not be
taught, not as the only way, but as an antidote to the prevailing
complacency? To show us that vagueness, subjectivity, and indeterminism,
are not forced on us by experimental facts, but by deliberate theoretical
choice? \cite[page 160]{Bell}
\eq

The last quoted sentence refers, of course, to the good news: that when we
consider, not the behavior of physicists but the physics itself, we find, in
the stark contrast between the claims of quantum philosophy and the actual
facts of quantum physics, compelling support for the objectivity and
rationality of nature herself.

Here is one more bit of information somewhat relevant in this regard. You
may well be wondering how, in fact, Bohm managed to accomplish what was so
widely regarded as impossible, and what his completion of quantum theory
involves. But you probably imagine that what eluded so many great minds
could not be conveyed in but a few minutes, even were this an audience of
experts.  However, the situation is quite otherwise. In order to arrive at
Bohmian mechanics from standard quantum theory one need do almost nothing!
One need only avoid quantum philosophy and complete the usual quantum
description in what is really the most obvious way: by simply including the
positions of the particles of a quantum system as part of the state
description of that system, allowing these positions to evolve in the most
natural way\cite{DGZ}.  The entire quantum formalism, including the
uncertainty principle and quantum randomness, emerges from an analysis of
this evolution (see \cite{DGZ,survey}). My long-time
collaborator, Detlef D\"urr\cite{dd}, has expressed this
succinctly---though in fact not succinctly enough---by declaring that the
essential innovation of Bohmian mechanics is the insight that {\it
particles move\/}! Bell, referring to the double-slit interference
experiment, put the matter this way:

\bq \noindent Is it not clear from the smallness of the scintillation on
the screen that we have to do with a particle? And is it not clear, from
the diffraction and interference patterns, that the motion of the particle
is directed by a wave? De Broglie showed in detail how the motion of a
particle, passing through just one of two holes in screen, could be
influenced by waves propagating through both holes.  And so influenced that
the particle does not go where the waves cancel out, but is attracted to
where they cooperate. This idea seems to me so natural and simple, to
resolve the wave-particle dilemma in such a clear and ordinary way, that it
is a great mystery to me that it was so generally ignored. \cite[page
191]{Bell} \eq
I think this should be a bit of a mystery for all of us!

I am grateful to Rebecca Goldstein and Eugene Speer for their comments and
suggestions. This work was supported in part by NSF Grants  DMS--9305930
and DMS--9504556.

\end{document}